\documentclass[useAMS,usenatbib]{mn2e}
\usepackage{epsfig}
\usepackage{amsmath}
\usepackage{amssymb}

\newcommand{\sinc}{{\rm sinc}}

\title{How achromatic is the stellar scintillation on large telescopes?}
\author[V.~Kornilov]{V.~Kornilov\thanks{E-mail: victor@sai.msu.ru}\\
Sternberg Astronomical Institute, Universitetsky prosp. 13, 119992 Moscow, Russia}

\begin{document}
\date{Accepted 2011 June 27.  Received 2011 June 26; in original form 2011 May 28}
\pagerange{\pageref{firstpage}--\pageref{lastpage}}
\pubyear{2011}
\maketitle

\label{firstpage}

\begin{abstract}
The atmospheric scintillation of stars is the main limitation of the accuracy of ground-based photometry of astronomical objects. This becomes particularly noticeable for a study of a variability with amplitudes on the order to thousandths of stellar magnitude or less. We examined the problem of the colour scintillation --- fluctuations of difference between light intensities measured simultaneously in two different photometric bands.

Relations between the colour scintillation power (index) and the atmospheric turbulence, telescope diameter and the characteristics of the photometric channels are derived. Asymptotic dependencies for large telescopes (1--10 m) are obtained that allow to predict the value of the colour scintillation for a particular telescope and detector. It is shown that the colour scintillation index is $\propto D^{-3}$ for measurements with both short (milliseconds) and long (seconds) exposures. The impact of the atmospheric dispersion which amplifies colour scintillation away from the zenith is estimated. We show that colour scintillation in the long-exposure regime depends strongly on the wind direction in the upper atmosphere.
\end{abstract}


\section{Introduction}

Photometric errors due to stellar scintillation have been investigated in a variety of works \citep[see e.g.][]{Young69,DravI,DravII}. A distinctive feature of these fluctuations is a very small angular correlation distance, on the order of the isoplanatic angle (typically a few arcseconds). Consequently, the use of a comparison star at larger angular distance only increases the scintillation errors. In some cases, it the scintillation, rather than photon statistics, defines the photometric accuracy \citep{Heasley1996,Everett2001}. The problem of the scintillation noise becomes very acute at integration times on the order of 1~s or less and on large telescopes.

A variety of methods were proposed for reducing the scintillation noise: use of multiple reference stars \citep{DravI,Gill1993}, choice of optimal spectral bands \citep{DravII}, arrays of small telescopes \citep{DravIII}, conjugation of the telescope entrance pupil to high turbulent layers \citep{Conj2010}, selection of optimal site for telescope installation \citep{Heasley1996,Kenyon2006}. This multitude of approaches demonstrates the relevance of the problem.

In case when the colour of an astronomical objects is variable, the use of an instrument which measures light intensity simultaneously in several photometric bands (channels) can significantly reduce the influence of the scintillation. Such approach was proposed in \citep{Fossat2003} to study intrinsic stellar oscillations \citep{Randall2005}. Various types of variability in close binary systems also often depend on the wavelength of the radiation \citep{Giov2001}.

It is commonly accepted that the stellar scintillation is achromatic in large telescope, i.e. when the diameter of the receiving aperture $D$ is much larger than the Fresnel radius of scintillation $r_F \sim 10\mbox{ cm}$. \citep{Roddier81}. However, this is true only asymptotically, but in actual situations there always remain some dependence on the wavelength $\lambda$.

The aim of this paper is to study the correlation of the scintillations in two spectrally different photometric bands. Typically, this correlation will be considered in terms of power of the differential signal --- differential {\it colour  scintillation}. A general description of the phenomenon is given in Sect.~\ref{sec:theory}.

In Sect.~\ref{sec:short_exp}, an asymptotic dependence of the power of colour scintillation on telescope diameter and altitude of the turbulence is studied. In this section, the case of short exposures is discussed, when smoothing of scintillation due to wind-driven turbulence motion is negligible. The case of significant averaging, i.e. long-exposure regime, is analysed in the following Section. In both regimes, the atmospheric dispersion greatly enhances the colour scintillation.

In the last Section, the effectiveness of multicolour technique (in the sense of simultaneous light intensity measurement in different photometric bands) for the scintillation noise suppression is evaluated.

\section{General formulae}
\label{sec:theory}

Basic expressions to calculate the covariance of light intensity fluctuations in two photometric bands are given in the paper \citep{multi2011}. They were obtained for the approximation of weak perturbations by the method previously used in \citep{polychrom2003}. The starting relation for the covariance $c_{1,2}$ of normalised intensities, arising due to turbulence in the layer located at the distance $z$ from the receiver, is the expression:
\begin{multline}
c_{1,2} = \langle I_1 I_2 \rangle - \langle I_1 \rangle\langle I_2 \rangle \approx \\
\approx 4\int\langle \bar\chi_1(\lambda_1)\bar\chi_2(\lambda_2)\rangle\,F_1(\lambda_1) F_2(\lambda_2)\,{\rm d}\lambda_1{\rm d}\lambda_2,
\label{eq:b12}
\end{multline}
where $\bar\chi_1$ and $\bar\chi_2$ are the averaged over entrance aperture logarithms of wave amplitudes in the first and the second channels, and $F_1(\lambda)$ and $F_2(\lambda)$ are normalised products of the photon sensitivity in the photometric bands with the energy distribution of the source spectrum. This equation is approximate because the right-hand side is written for the log-amplitudes and the left-hand side for the intensities. For weak scintillation, the expression can be regarded as accurate.

Using the formulae 7--11 from the paper \citep{polychrom2003}, one can calculate the covariance of the log-amplitudes $\langle\bar\chi_1(\lambda_1)\bar\chi_2(\lambda_2)\rangle$ where apertures of the photometric channels may be not be coincident, in general. For coincident circular apertures, the product of the Fourier transforms of the aperture functions $\tilde W_1\tilde W_2$ is replaced by the axi-symmetric aperture filter $A(f)$ during further integration over the two-dimension spatial frequency.

\subsection{Correlation of the scintillation in two photometric bands}
\label{sec:theory-corr}

The relationship between the scintillation power ({\it scintillation  index}) $s^2$ and the turbulence  intensity $C_n^2(z)$ at the distance $z$ is described with help of {\it weighting functions} (WF), that is scintillation power produced by a turbulent layer of unity intensity. In the linear approximation of the weak perturbations
\begin{equation}
s^2 = \int_A C_n^2(z)\, Q(z) {\rm d}z.
\end{equation}
where $Q(z)$ is the respective WF and the integration is performed over whole atmospheric depth. In analogy with the expression for $Q(z)$ in \citep{polychrom2003}, we can obtain an expression for the {\it covariance weighting function} $R_{1,2}(z)$ which describes the covariance of scintillation in two different photometric channels:
\begin{equation}
R_{1,2}(z) = 9.61\int_0^\infty f^{-8/3}S_{1,2}(z,f) A(f) \,{\rm d}f,
\label{eq:r12}
\end{equation}
where the integration is performed over the modulus the spatial frequency $f$, $A(f)$ is the spectral aperture filter, equal to $A(f) = (2J_1(\pi Df)/(\pi Df))^2$ for the case of coincident circular apertures of the diameter $D$, and the kernel $S_{1,2}(z,f)$ is the {\it Fresnel spectral filter}. This filter is a double integral which splits into a product of two integrals:
\begin{multline}
S_{1,2}(z,f) = \int_{F_1} \frac{F_1(\lambda_1)}{\lambda_1}\sin(\pi\lambda_1 z f^2) \,{\rm d}\lambda_1 \\
\times \int_{F_2} \frac{F_2(\lambda_2)}{\lambda_2} \sin(\pi\lambda_2 z f^2) \,{\rm d}\lambda_2.
\label{eq:s12}
\end{multline}

Then, the observed covariance $c_{1,2}$ of the scintillation in two channels ({\it cross-index}) depends on the vertical distribution of the structure coefficient $C_n^2(z)$:
\begin{equation}
c_{1,2} = \int_A C_n^2(z) \, R_{12}(z) \,{\rm d}z.
\end{equation}

If $D \gg r_F = (\lambda z )^{1/2}$), we can use the method of \citep{Roddier81}, putting $\sin(\pi\lambda zf^2) \approx \pi\lambda zf^2$. The asymptotic dependence for the covariance WF $R_{1,2} (z) = 17.33\,D^{-7/3}z^2$ is obtained. This expression is identical to the known relation for the scintillation power in the case of large telescopes.

\subsection{Colour scintillation}

Correlation of the scintillation can be described in terms of the {\it differential colour index} --- the variance of the difference of normalised signals in two photometric channels, an analogue of differential spatial index \citep{timeconst2002}. In the case of strong correlation expected for large telescopes, this approach is more practical for measurement and analysis. This colour scintillation arises from the dependence of the diffraction on the wavelength of light. Based on the relation for the variance of a difference:
\begin{equation}
s^2_d = s^2_1 + s^2_2 - 2\,c_{1,2},
\label{eq:s2_diff}
\end{equation}
the colour weighting function $Q_{d}(z)$ can be represented as:
\begin{equation}
Q_{d}(z) = Q_{1}(z) + Q_{2}(z) - 2 R_{1,2}(z),
\label{eq:weight_diff}
\end{equation}
Returning to the expression (\ref{eq:r12}) and summing the corresponding integrands, we obtain
\begin{equation}
Q_{d}(z) = 9.61\int_0^\infty f^{-8/3}T_{1,2}(z,f) A(f) \,{\rm d}f,
\label{eq:weight_qd}
\end{equation}
where $T_{1,2}(z,f)$ is the {\it differential Fresnel filter}: 
\begin{multline}
T_{1,2}(z,f) = \iint\limits_{F_1,F_2}{\rm d}\lambda_2 \,{\rm d}\lambda_1 \\
\times \Bigl(\frac{F_1(\lambda_1)}{\lambda_1}\sin(\pi\lambda_1 z f^2) - \frac{F_2(\lambda_2)}{\lambda_2} \sin(\pi\lambda_2 z f^2)\Bigl)^2.
\label{eq:t12}
\end{multline}
If the usual Fresnel filter used for the $Q(z)$ calculation has behaviour $\propto f^4$ near $F=0$, the differential Fresnel filter  $T(z,f)$ is proportional to $f^{12}$ there and suppresses the low frequencies passed by the aperture filter very effectively.

\begin{figure}
\centering
\psfig{figure=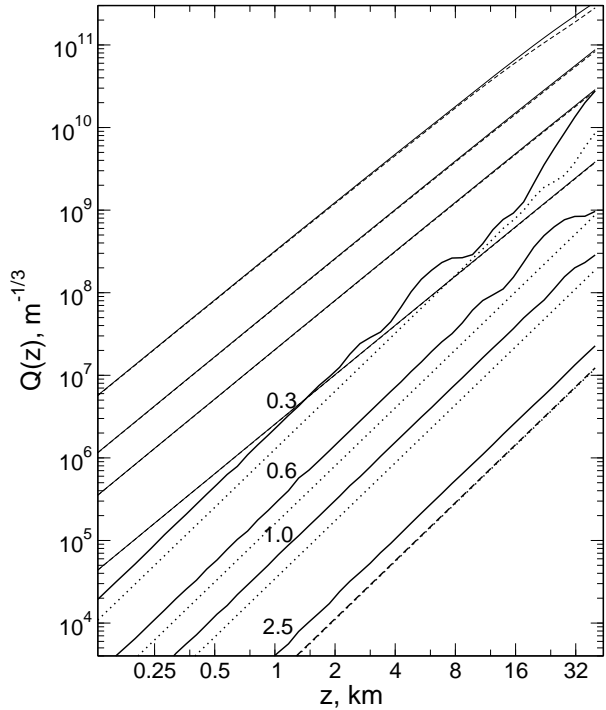,height=10.5cm}
\caption{ WFs for the 0.3~m, 0.6~m, 1.0~m and 2.5~m telescopes with  central obscuration 0.3 in photometric bands $B$ (thin solid line)  and $R$ (thin dashed). The colour WFs $Q_{B,R}(z)$ are shown with a  thick line. The cases of the apertures without central obscuration  are shown by dotted lines. The thick dashed line represents the  estimate calculated with the formula (\ref{eq:assimp}) \label{fig:wfs_large}}
\end{figure}

For apertures of arbitrary size and wide photometric bands, the functions $R_{1,2}(z)$ and $Q_{d}(z)$ can be obtained by numerical integration with formulae (\ref{eq:r12}), (\ref{eq:s12}) and (\ref{eq:weight_qd}), (\ref{eq:t12}), as done when calculating the WFs in the program {\sl atmos}  developed for processing of measurements with Multi-Aperture Scintillation Sensor (MASS) \citep{MASS,mnras2003}. The computed functions $Q_{d}(z)$ for the $B$ and $R$ photometric bands (effective wavelengths $\lambda_B = 440$~nm É $\lambda_R = 690$~nm) and for four different telescope apertures are shown in Fig.~\ref{fig:wfs_large}.

\section{Short exposures}
\label{sec:short_exp}
\label{sec:zero_exp}

In the case of large apertures $D \gg (\lambda z)^{1/2}$, the asymptotic dependencies for both index and cross-index are completely achromatic (\ref{sec:theory-corr}). Therefore, it is not possible to obtain the asymptote of the colour index $s_d^2$ from the formula (\ref{eq:s2_diff}). However, the asymptote can be found from the analysis of the colour WF (\ref{eq:weight_qd}).

Writing this formula for two monochromatic channels $\lambda_1$ and $\lambda_2$ and a circular aperture of diameter $D$ and passing to the dimensionless frequency $q = fD$, we get
\begin{multline}
Q_{d}(z) = 9.61\, D^{5/3}\int_0^\infty \left[\frac{2J_1(\pi q)}{\pi q} \right]^2 q^{-8/3} \\ \times \left(\frac{\sin(\pi\lambda_1 z q^2/D^2)}{\lambda_1} - \frac{\sin(\pi\lambda_2 z q^2/D^2)}{\lambda_2}\right)^2\, {\rm d}q.
\label{eq:wdz}
\end{multline}

\begin{figure}
\psfig{figure=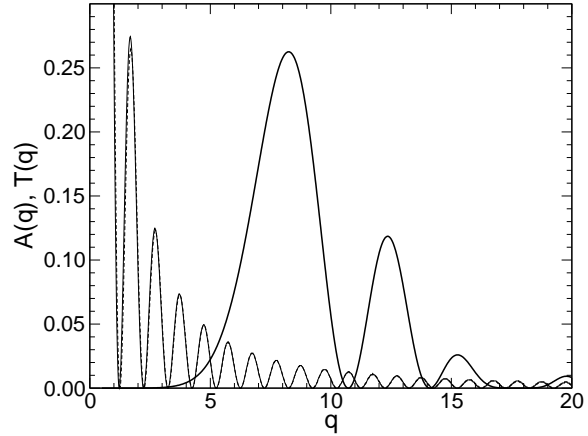,height=8.2cm,angle=-90}
\caption{Position and typical form of the aperture $A(q)$ (thin  line) and differential Fresnel $T(q)$ (thick line) spectral  filters. The approximation of the aperture filter is shown by the dashed  line.
\label{fig:filters}}
\end{figure}

\begin{figure}
\psfig{figure=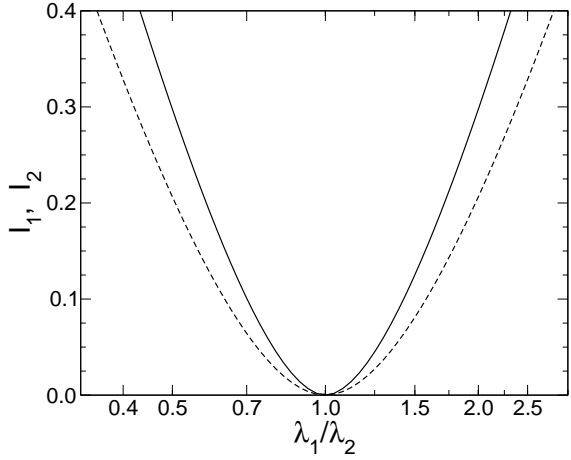,height=8.2cm,angle=-90}
\caption{Integrals $\mathcal I_1$ in the  expression \ref{eq:assimp} (solid line) and $\mathcal I_2$ in the  expression \ref{eq:assimp_w} as a function of the parameter $a^2 =  \lambda_1/\lambda_2$.
\label{fig:integrs}}
\end{figure}

Clearly, the integral still depends on the parameters $z$, $D$, $\lambda_1$ and $\lambda_2$. In this case, the aperture $A(q)$ and the differential $T(q)$ filters almost do not overlap (Fig.~\ref{fig:filters}). In the region where $T(q)$ is significantly different from zero, one can use the asymptotic form of the aperture filter:
\begin{equation}
\left[\frac{2J_1(\pi q)}{\pi q} \right]^2 = \frac{8}{\pi^4 q^3}\cos^2(\pi q - \frac{3}{4}\pi),
\label{eq:j1_approx}
\end{equation}
which is valid when $q \gg 1$. This approximation is shown in Fig.~\ref{fig:filters} and one can see that in the area of interest, it coincides with the exact function $A(q)$. Since $A(q)$ oscillates within the envelope of $T(q)$, it is possible to replace $\cos^2$ with its average value 1/2:
\begin{multline}
Q_{d}(z) = 9.61\, D^{5/3} \frac{4}{\pi^4} \int_0^\infty q^{-17/3} \\
\times \left(\frac{\sin(\pi\lambda_1 z q^2/D^2)}{\lambda_1} - \frac{\sin(\pi\lambda_2 z q^2/D^2)}{\lambda_2}\right)^2\, {\rm d}q,
\end{multline}
This operation allows us to reduce the integrand to a dimensionless form by replacing the variable $u = \pi(\lambda_1\lambda_2)^{1/2} z
D^{-2} q^2$:
\begin{multline}
Q_{d}(z) = 2.85\, z^{7/3} D^{-3}\,(\lambda_1 \lambda_2)^{1/6} \\
\times \int_0^\infty u^{-10/3}\, \left(\frac{1}{a}\sin(a u) - a \sin(u/a)\right)^2\, {\rm d}u,
\label{eq:assimp}
\end{multline}
where $a = (\lambda_1/\lambda_2)^{1/2}$. The integral $\mathcal I_1$ in this formula can not be represented as a sum of three integrals, because each separate component diverges. We could not calculate $\mathcal I_1$ analytically, therefore its behaviour as a function of the ratio $\lambda_1/\lambda_2 = a^2$ was calculated numerically and is presented in Fig.~\ref{fig:integrs}. Near the point of $\lambda_1 = \lambda_2$ the dependence is quadratic, both branches increase indefinitely at small and large arguments.

\subsection{Analysis of the asymptotic dependence}
\label{sec:analys}

The obtained asymptotic dependence (\ref{eq:assimp}) is confirmed by the numerical integration of the formula (\ref{eq:t12}). The polychromatic WFs for 0.3~m, 0.6~m, 1.0~m and 2.5~m apertures and $B$ and $R$ photometric bands are shown in Fig.~\ref{fig:wfs_large}. Complex structure of $T(q)$ affects the form of the WFs, they have a noticeable waves with altitude. The figure also shows the WFs for the same aperture without a central obscuration for comparison with the asymptotic estimates (\ref{eq:assimp}). In the scale of the figure, the both curves coincide to within 1--3~percent.

Such a good match between the polychromatic and the monochromatic curves occurs because the scintillation is only slightly chromatic in the case of a large aperture. As a consequence, the functions $Q_{d}(z)$ are virtually independent of the spectral band width. This means that estimates based on the formula (\ref{eq:assimp}) can be used for real photometric bands. To account for the central obscuration 0.3 (typical for a two-mirror telescope), it is sufficient to multiply the asymptotic estimate by the factor $1.9$. For the normal scintillation, this factor is less than $\approx 1.21$.

Evident relation between the altitude $h$ and the distance $z = hM_z$ predicts that the colour scintillation power is a function on the air mass $M_z$ in the form of $s^2_d \propto M_z^{10/3}$ due to simple geometric factor of light propagation.

The dependence of colour scintillation on altitude is stronger than for the conventional scintillation. It means that the colour scintillation is generated at the higher layers of the turbulent atmosphere. Given that the scintillation index for the case of short exposure is defined by the known relation \citep{Roddier81}:
\begin{equation}
s^2 = 17.33\, D^{-7/3}\int_A C_n^2(z)\,z^2 {\rm\,d}z,
\end{equation}
and that the colour index is equal to
\begin{equation}
s^2_d = 2.852\, D^{-3} (\lambda_1\lambda_2)^{1/6} {\mathcal I_1(\lambda_1/\lambda_2)}\int_A C_n^2(z)\,z^{7/3} {\rm\,d}z,
\end{equation}
it is possible to estimate the relative power of $s^2_d/s^2$ --- {\it  achromatic factor} of the scintillation. For that, we take $z^{1/3}$ out the integral, replacing it with average value $\langle{z^{1/3}}\rangle$ weighted with $C_n^2(z)\,z^2$. Then the ratio of the indices can be written as
\begin{equation}
\frac{s^2_d}{s^2} = 0.165\, D^{-2/3}(\lambda_1\lambda_2)^{1/6} \langle{z^{1/3}}\rangle {\mathcal I_1(\lambda_1/\lambda_2)}.
\label{eq:ahrom} 
\end{equation}
Estimates of the altitude atmospheric moments $M_p = \int C_n^2(z)\,z^p {\rm \,d}z$ can be obtained from vertical distributions of the optical turbulence measured with the MASS instrument \citep{MASS}. For example, measurements at Mount Shadzhatmaz in 2007--2009 show the median of $M_2 = 1.1\cdot10^{-5}\mbox{ m}^{7/3}$ \citep{kgo2010}. The quantity $\langle{z^{1/3}}\rangle$ depends only weakly on the actual turbulence profile and can be set equal to $24\mbox{ m}^{1/3}$. This assumption ensures accuracy of the estimate not worse than 10~percent.

The expression (\ref{eq:ahrom}) shows that the achromatization comes not very quickly: increasing the telescope diameter by 2.5~times reduces the chromaticity only by half (the scintillation amplitude is reduced by 1.5~times). For small apertures (on the order to the Fresnel radius), the colour scintillation is $\sim 0.1$ of the normal one (or $\sim 0.3$ in amplitude). In the case of 2.5~m telescope, the achromatic factor is $\sim 0.003$ in power or $\sim 0.05$ in amplitude when measurements are performed in the $B$ and $R$ bands.

\subsection{Decorrelation due to atmospheric dispersion}
\label{sec:ad}

When measurements are done away from the zenith, beams falling into the first $\lambda_1$ and the second $\lambda_2 $ channels, pass through atmosphere by a few different ways because of the effect of atmospheric dispersion. That leads to a decorrelation of the scintillation, i.e. to an increase of the colour scintillation. This phenomenon was discussed in \citep{DravII,Hubb81,caccia1988}. In the paper \citep{multi2011}, the expression for the additional power $\Delta s^2_d$ of the colour index $s^2_d$ was presented. The WF for calculating the extra power due to atmospheric dispersion is described by the formula:
\begin{equation}
Q_{\Delta}(z) = 9.61\int\limits_0^\infty f^{-8/3} S_{1,2}(z,f)\,(2-2J_0(2\pi xf))A(f)\,{\rm d}f,
\label{eq:ddz}
\end{equation}
where $A(f)$ is the aperture filter as usual, and $S_{1,2}(z,f)$ is the Fresnel filter of the kind of (\ref{eq:s12}), and $x = x(z)$ is the displacement of one beam with respect to the other at the distance $z$.

Recall that for a flat isothermal atmosphere, the beam displacement $x$ at the altitude $h$ is given by:
\begin{equation}
x = \tan\gamma\,\sec\gamma\,\Delta n_{1,2}\,e^{-h_o/H_0}\bigl( H_0-(H_0+h)e^{-h/H_0}\bigr),
\label{eq:x-z}
\end{equation}
where $\gamma$ is the zenith distance, $h_o$ is the observatory level above sea, $h = z\,\cos\gamma$ is the height of turbulent layer with respect to the telescope level, $H_0$ is the atmospheric height scale ($\approx 8300$~m). The multiplier can be written also as $\tan\gamma\,\sec\gamma = M_z\,(M_z^2-1)^{1/2}$ if the air mass $M_z$ is used instead of $\gamma$. We emphasise that the $x$ is not a horizontal shift, but displacement perpendicular to the light direction. With height increasing, $x$ grows approximately quadratically and at altitude about 14~km ($1.68\,H_0$) it reaches 1/2 of its maximum value, then asymptotically approaches it.

It is convenient to estimate the effect when $\tan\gamma\,\sec\gamma = 1$ or air mass $M_z = 1.27202 = \sqrt{\phi}$ ($\phi$ is the golden ratio). The shift outside the atmosphere is $x_{\infty} = H_0\,\Delta n_{1,2}\,e^{-h_o/H_0}$. For the $B$ and $R$ photometric bands and observatory at 2500~m, the difference between the refractive indices at the sea level is $\Delta n_ {B, R} = -5.15\cdot 10^{-6}$ and $x_{\infty} = 0.032$~m.

\begin{figure}
\centering
\psfig{figure=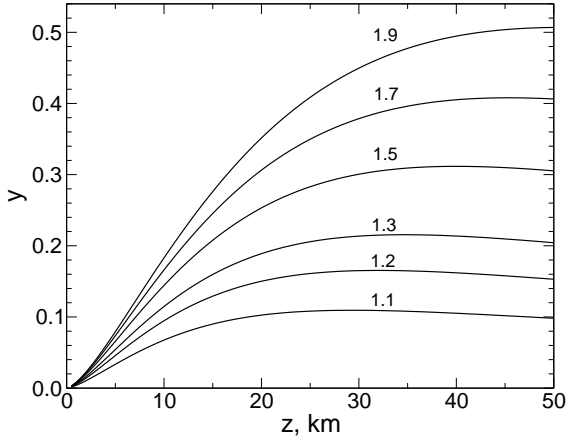,height=8.2cm,angle=-90}
\caption{Relative displacement of the beams $ y = x/r_{F}$  versus distance to the turbulent layer for different air masses  (labels on the curves) for the $B$ and $R$ photometric bands.\label{fig:dispace}}
\end{figure}

Let us investigate the $Q_{\Delta}(z)$ behaviour in the case of two monochromatic channels and circular telescope aperture:
\begin{multline}
Q_{\Delta}(z) = 9.61\,\int_0^\infty f^{-8/3}\, \frac{\sin(\pi\lambda_1  z f^2)}{\lambda_1} \frac{\sin(\pi\lambda_2 z  f^2)}{\lambda_2}\\
\times (2-2\,J_0(2\pi xf))\left(\frac{2J_1(\pi Df)}{\pi  Df}\right)^2 \,{\rm d}f.
\label{eq:wddz}
\end{multline}
One can introduce the dimensionless frequency $q = r_Ff$ and the relative shift $y = x/r_F$. Calculation with (\ref{eq:x-z}) shows that for all reasonable air masses the condition $y \la 1$ holds because $x$ increases with altitude and $r_F$ increases with distance to the layer. The dependence of $y$ on the distance is shown in Fig.~\ref{fig:dispace} for different air masses. For $B$ and $R$ photometric bands, the maximum of $y$ is related to the air mass as $y = 0.22\,M_z^{1/2}\,(M_z^2-1)^{1/2}$. The average compound Fresnel radius $r_F = z^{1/2}(\lambda_1\lambda_2)^{1/4}$ is used here. The main peak of the integrand lies in the domain $q \la 1$. Note that the Fresnel cross-filter for some frequencies may be negative.

We replace the aperture filter $A(f)$ with its approximation (\ref{eq:j1_approx}) which is applicable when $q > r_F/\pi D$. Just as before, we can replace $\cos^2$ with $1/2$. Unlike the case considered in Sect.~\ref{sec:zero_exp}, the cross-filter is proportional to $\sim q^4$ near zero and it does not provide effective suppression of the approximation error. Thus, the approximate integrand has an infinite derivative at $q = 0$, while the exact expression has zero derivative. Such distinction leads to $\approx $10~percent error, which was estimated from a comparison of the asymptote with numerical integration of the formula (\ref{eq:wddz}).

\begin{figure}
\centering
\psfig{figure=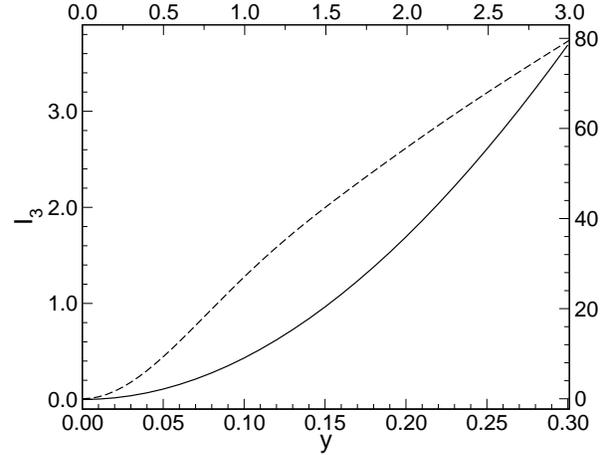,height=8.2cm,angle=-90}
\caption{Dependence of the integral ${\mathcal I_3}$ on the relative  displacement. The initial part of the curve is shown with the solid  line with coordinate axes on the left and below. The total  dependence is shown with dashed line with axes at the top and  right. \label{fig:i3_y} }
\end{figure}

As a result of these transformations, the expression reduces to the dimensionless integral ${\mathcal I_3(a,y)}$, where $a = (\lambda_1/\lambda_2)^{1/2}$, with the dimensions factor in front of it:
\begin{multline}
Q_{\Delta}(z) = 0.789\, D^{-3}z^{7/3}(\lambda_1\lambda_2)^{1/6}\\
\times \int_0^\infty q^{-17/3}\,
\sin(\pi a q^2)\sin(\pi q^2/a)(1-J_0(2\pi y q))\,{\rm d}q.
\label{eq:qdelta}
\end{multline}
The ${\mathcal I_3}$ depends implicitly on the distance to the layer $z$ through $y$, but it is not very important for a general estimate.

The dependence on $y$ can be obtained by replacing the filter $1-J_0(2\pi yq)$ with its approximate expression $\pi^2 y^2 q^2$ which is applicable when $y \la 1$. In this case, the integral
\begin{equation}
{\mathcal I_3} \approx 17.3\,y^2(\lambda_1\lambda_2)^{-4/6}\bigl( (\lambda_1+\lambda_2)^{4/3} - (\lambda_1-\lambda_2)^{4/3} \bigr).
\label{eq:i3appr}
\end{equation}
The relative displacement $y$ depends implicitly on the wavelength ratio $a$ through the refractive index difference (\ref{eq:x-z}). As the difference between the wavelengths gets larger, the factor depending on $a$ in (\ref{eq:i3appr}) becomes smaller, but this is compensated by the increased $y^2$.

In the case $\lambda_1 = \lambda_2$, the integral ${\mathcal I_3}$ is calculated analytically after differentiation over $y$, but in a rather complicated way through hypergeometric functions. Its behaviour is shown in Fig.~\ref{fig:i3_y} in two scales. We see that for $y < 0.3$ the dependence is quadratic, then it progressively becomes weaker than linear. Comparison with the formula (\ref{eq:i3appr}) demonstrates that the latter provides estimates with an accuracy of about 5 percent up to $y = 0.3$. Consequently, for air masses $M_z \la 1.5$, the use of quadratic dependence of $\Delta s^2_d \propto y^2 $ or $\Delta s^2_d \propto M_z\,(M_z^2-1)$ is sufficient for practical purposes.

\subsection{Relative contribution of the additional colour scintillation}

Comparison of the formula (\ref{eq:qdelta}) with the (\ref{eq:assimp}) for the colour scintillation shows that their ratio depends only on the wavelengths of the photometric channels and on the relative displacement $y$. The relative contribution of the atmospheric dispersion to colour scintillation will be determined as follows:
\begin{equation}
\frac{\Delta s^2_d}{s^2_d} = 0.277\,\frac{\mathcal I_3(y,\lambda_1/\lambda_2))}{\mathcal I_1(\lambda_1/\lambda_2)}.
\label{eq:rel_ampl}
\end{equation}
Using the plots in Fig.~\ref{fig:integrs} and Fig.~\ref{fig:dispace} one can estimate that the ratio $\Delta s^2_d/s^2_d$ reaches 1 at $y \approx 0.1$. Such value corresponds to observations at air mass $M_z \approx 1.1$. Consequently, colour scintillation caused by the propagation will dominate over colour scintillation caused by the atmospheric dispersion only near the zenith ($\gamma < 25^\circ$), when measured in the $B$ and $R$ bands.

\begin{figure}
\centering
\psfig{figure=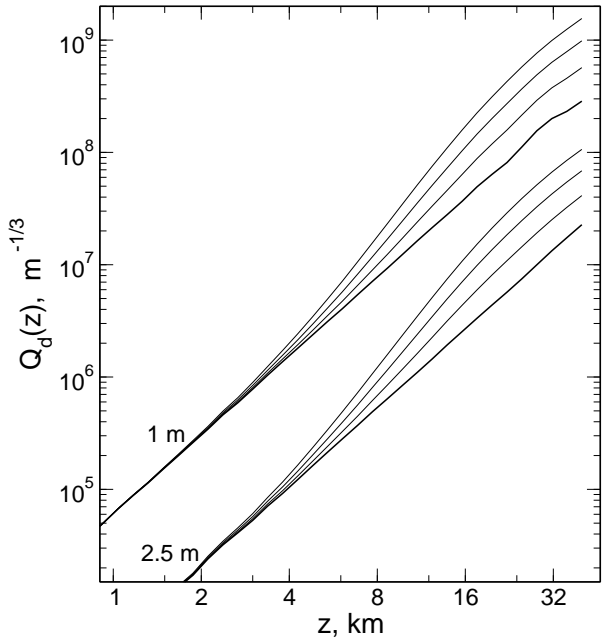,height=9.6cm}
\caption{Colour WFs for 1~m and 2.5~m telescopes. The functions  $Q_{d}(z) + Q_{\Delta}(z)$ for short exposures at air mass 1.0  (thick lines), 1.1, 1.2 and 1.3 (bottom to top).
\label{fig:real_disp_s}}
\end{figure}

We improve the expression (\ref{eq:ahrom}) by adding extra colour scintillation induced by the atmospheric dispersion:
\begin{equation}
\frac{s^2_d+\Delta s^2_d}{s^2} \approx (0.005+0.021\,M_z(M_z^2-1))\,D^{-2/3}.
\label{eq:final0}
\end{equation}
Numerical integration of the initial expression taking in account all altitudinal dependencies, leads to the WFs which are shown in Fig.\ref{fig:real_disp_s}, where a family of functions $Q_{d}(z)+Q_{\Delta}(z)$ is plotted for 1~m and 2.5~m telescopes and air masses 1.0, 1.1, 1.2 and 1.3.

\section{Long exposures}
\label{sec:long_exp}

In the previous sections, the case of measurements with temporal scales comparable to the atmospheric coherence time was considered. Conventional photometry rarely has to deal with such short exposures and therefore, the study of the colour scintillation on time scales of seconds and more has a larger practical interest. As a rule, on such time scales the {\it long exposure regime} is realised \citep{wind2010} when the turbulence drifts during exposure time $\tau$ at the distance $w\tau \gg D$ (here $w$ is the wind speed). The scintillation power determines the photometric errors in this case.

It was shown in \citep{multi2011} for the case of an infinitely small aperture that the wind averaging increases the correlation of scintillations, i.e. reduces the colour scintillation. Naturally, in that case the condition $w\tau \gg r_F$ was used as a criterion of the long-exposure regime.

\subsection{Wind smoothing}
\label{sec:wind_zero}

Using the approach based on the hypothesis of frozen turbulence and described in \citep{wind2010}, we modify suitably the expression (\ref{eq:wdz}) by adding the spectral filter ${\mathcal T}_1(w\tau f)$ of the wind shear, equal to $1/\pi w(z) \tau f$ in the long-exposure regime. Note that $w(z)$ is the component of the wind vector transverse to the line of sight. Omitting the intermediate transformations, we write the analogue of (\ref{eq:assimp}) for the colour WF:
\begin{multline}
U^{\prime}_{d}(z) = 1.609\, z^{17/6} D^{-3}\,(\lambda_1 \lambda_2)^{5/12}\\
\times \int_0^\infty u^{-23/6}\, \left(\frac{1}{a}\sin(a u) - a \sin(u/a)\right)^2\, {\rm d}u,
\label{eq:assimp_w}
\end{multline}
The dimension of this function is $m^{2/3}$ because for calculation of the scintillation index $\tilde s^2_d$, it should be divided by the wind shear $w(z)\tau$. The integral $\mathcal I_2$ in this formula can be calculated numerically. Its behaviour depending on the ratio $\lambda_1/\lambda_2 = a^2$, is shown in Fig.~\ref{fig:integrs}. The behaviour is not fundamentally different from that of $\mathcal I_1$ in (\ref{eq:assimp}).

Attention is drawn to an almost cubic dependence on the altitude. This means that for long exposures a stratospheric turbulence at altitudes of 25--30~km will contribute to the colour scintillation. Numerical integration of the initial expression confirms the asymptote in the form of (\ref{eq:assimp_w}).

In the long-exposure case, the scintillation becomes more achromatic with increasing $D$ faster than at short exposures. Indeed, using the asymptote for $U^{\prime}(z) = 10.66\,D^{-4/3}\,z^2$ from the paper \citep{wind2010}, the following expression can be written (the scintillation indices for long exposures are denoted as $\tilde s^2_d$ and $\tilde s^2$):
\begin{equation}
\frac{\tilde s^2_d}{\tilde s^2} = 0.151\, D^{-5/3}(\lambda_1\lambda_2)^{5/12} \langle{z^{5/6}}\rangle {\mathcal I_2(\lambda_1/\lambda_2)}.
\label{eq:ahrom_w}
\end{equation}
The quantity $\langle{z^{5/6}}\rangle$ is the $z^{5/6}$ averaged with the weight $C_n^2(z)\,z^2/w(z)$. The wind velocity dependence on altitude $w(z)$ additionally increases the effective altitude of colour scintillation above the tropopause, because a decrease of turbulence intensity with height is offset by a decrease in the wind speed above the tropopause. Based on the actual vertical distributions of turbulence (see Sect.~\ref{sec:analys}) we can set $\langle{z^{5/6}}\rangle \approx 3300\mbox{ m}^{5/6}$. This estimate has uncertainty about 20~percent. For 2.5~m telescope the $\tilde s^2_d/\tilde s^2$ is equal to $0.0001$ in power or $0.01$ in amplitude, which is considerably less than in the case of short exposures.

The dependence of the colour scintillation on the air mass $M_z$ is almost the fourth degree and when the wind blows along line of sight, becomes $\propto M_z^{29/6}$ because the $\tilde s_d^2$ is inversely proportional to the transversal wind which decreases as $M_z^{-1}$. The achromatic factor (\ref{eq:ahrom_w}) increases as $M_z^{5/6}$ at all times.

\subsection{Decorrelation due to atmospheric dispersion}

In the long exposure regime, additional colour scintillation due to atmospheric dispersion will depend on the angle $\theta$ between the direction of the wind vector and the azimuth of the star. Indeed, during long exposure the wavefront is averaged along the wind direction over an area of about $w\tau D$. When the displacement $x$ is transverse to the wind, the non-overlapping area is $\approx 2xw\tau$, much larger than $\approx 2xD$ for the case when $x$ is parallel to the wind.

Therefore, during the transition from two-dimensional integration over spatial frequency to the integration over its modulus, we must average over the angle $\phi$ the product of the two filters, the one for aperture displacement and another for the wind shear. In the equation (\ref{eq:ddz}) the factor of $2-2\,J_0(2 \pi xf)$ is the result of the averaging over the angle of the asymmetric filter of the aperture displacement $2-2\cos(2\pi xf\cos\phi) = 4\sin^2(\pi xf\cos\phi)$. We include the filter of the wind shear in the form of $\sinc^2(w\tau f\cos\phi)$ by rotating the coordinate x-axis by the angle $\theta$ in the direction of the wind. The product is averaged over $\phi$:
\begin{equation}
\mathcal L = \frac{1}{2\pi}\int_0^{2\pi} 4\sin^2(\pi xf\cos(\phi-\theta))\,\sinc^2(w\tau f\cos\phi){\,\rm d}\phi.
\end{equation}
The asymptote of the integral $\mathcal L$ at $w\tau \to \infty $ depends on the angle $\theta$. Since $xf = yq \la 1$, the linear approximation $\sin^2(\pi xf\cos(\phi-\theta)) \approx \pi^2x^2f^2\cos^2(\phi-\theta)$ may be used. For $\theta = 0$ (same wind and source azimuth), this gives:
\begin{equation}
\mathcal L_{\parallel} = \frac{x^2}{2\,w^2\tau^2}(1-J_0(4\pi w\tau f)).
\end{equation}
If $\theta = \pi/2$ (wind direction perpendicular to the atmospheric dispersion), the averaging results in:
\begin{multline}
\mathcal L_{\perp} = 4\pi^2x^2f^2\,{\mathcal T}_1(w\tau f) - \frac{x^2}{2\,w^2\tau^2}(1-J_0(4\pi w \tau f)) \\
= 4\pi\frac{x^2\,f}{w\tau} - \mathcal L_{\parallel},
\end{multline}
where ${\mathcal T}_1(w\tau f)$ is the wind filter \citep{wind2010} which equals $1/\pi w\tau f$ in the long-exposure regime.

Comparison of the relations for $\mathcal L_{\perp}$ and $\mathcal L_{\parallel}$ shows that both components are approximately equal at very low frequencies $f \la 1/\pi w\tau$. At larger frequencies, $\mathcal L_{\perp} \gg \mathcal L_{\parallel}$. Taking into account that the maximum of the scintillation is at $f = 0.555/r_F \sim 5-10$ (maximum of the product $q^{-8/3}\,\sin^2(\pi q^2)$) and that $1/\pi w\tau \ll 0.1$, it can be argued that the investigated effect is mainly determined by the perpendicular component of wind.

Corresponding WF $U^{\prime}_{\Delta}(z)$ can be obtained by replacing the factor $2-2\,J_0(2\pi yq)$ with the expression for $\mathcal L$ in the formula (\ref{eq:ddz}) or in formula (\ref{eq:wddz}) in the case of two monochromatic bands and a circular aperture. To get a formula like (\ref{eq:assimp_w}), we have to isolate the factor $1/w\tau$.

Except for the factor $1-J_0(4\pi w\tau f) $ which almost everywhere is equal to unity, the component $\mathcal L_{\parallel}$ does not depend on the frequency and can be taken out of the integral. In this case
\begin{multline}
U^{\prime}_{\Delta}(z)_{\parallel} = \frac{x^2}{2\,w\tau}\,R_{1,2}(z) = 8.66\frac{x^2}{w\tau} D^{-7/3}z^2 \\
= 8.66\frac{y^2}{w\tau} D^{-7/3}z^3(\lambda_1\lambda_2)^{1/2},
\label{eq:uddz_par}
\end{multline}
where $R_{1,2}(z)$ is the cross-index WF (\ref{eq:r12}) having the usual asymptote.

In calculating the component $\mathcal L_{\perp}$, one can neglect the term $\mathcal L_{\parallel}$. After replacing $f$ with the dimensionless frequency $q$, replacing the aperture filter with its approximation (\ref{eq:j1_approx}), separation of $1/w\tau$ and easy transformations we obtain:
\begin{multline}
U^{\prime}_{\Delta}(z)_{\perp} = 4.96\,y^2\,D^{-3}(\lambda_1\lambda_2)^{5/12}z^{17/6}\\
\times \int_0^\infty q^{-14/3}\, \sin(\pi a q^2)\sin(\pi q^2/a)\,{\rm d}q.
\label{eq:uddz_per}
\end{multline}
The approximation of the integrand near zero frequency is poor because the power of the $q^{-14/3}$ term is close to the limit for integrability. Numerical integration of the original formula shows that the error of the asymptotic expression (\ref{eq:uddz_per}) is on the order 20~percent for $D = 2-4\mbox{ m}$ and $z \approx 4\mbox{ km}$. In this equation, the integral can be calculated analytically and it equals to:
\begin{equation}
{\mathcal I_4^{\prime}} = 7.175\,(\lambda_1\lambda_2)^{-11/12}\bigl( (\lambda_1+\lambda_2)^{11/6} - |\lambda_1-\lambda_2|^{11/6} \bigr).
\end{equation}

\begin{figure}
\centering
\psfig{figure=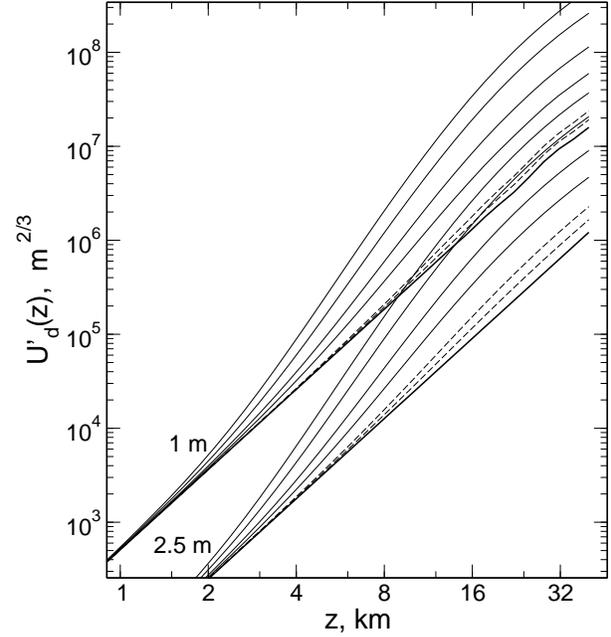,height=9.6cm}
\caption{Colour WFs for 1~m and 2.5~m telescopes.  The functions  $U^{\prime}_{d}(z) + U^{\prime}_{\Delta}(z)$ for measurements with  long exposures; the thick line corresponds to observations at  zenith, thin solid lines --- at the air masses 1.05, 1.1, 1.2 and  1.3 for the transversal wind and the thin dashed lines --- at air  masses 1.3 and 1.5 for the longitudinal wind.
\label{fig:real_disp_l}}
\end{figure}

\subsection{Relative contribution of the wind shear to colour scintillation}

The formulae (\ref{eq:uddz_par}) and (\ref{eq:uddz_per}) allow us to calculate the relative increase of the colour scintillation caused by the turbulence wind drift. To do this, we can use the method of Sects.~\ref{sec:analys} and \ref{sec:wind_zero}. In the case of parallel wind we have:
\begin{equation}
\frac{\Delta\tilde s^2_d}{\tilde s^2_d} = 5.38\,\frac{y^2}{w\tau}\frac{(\lambda_1\lambda_2)^{1/12}} {\mathcal I_2(\lambda_1/\lambda_2)}D^{2/3}\langle{z^{1/6}}\rangle,
\end{equation}
where $\langle{z^{1/6}}\rangle$ is the average of $z^{1/6}$ with the weight $C^2_n(z)\,z^{17/6}/w(z)$. This weight has maximum at even higher altitudes than in the case of (\ref{eq:ahrom_w}); for the typical conditions considered here we estimate $\langle{z^{1/6}}\rangle \approx 5\mbox{ m}^{1/6}$. For the $B$ and $R$ bands we can use the approximation $\approx 30\,y^2 D^{2/3}/{w\tau}$ which gives the magnification factor $\la 2$ for $ y = 0.3$ ($M_z \approx 1.5$), wind shear $w\tau = 5$, and 1~m telescope.

The proportionality to $D^{2/3}$ looks strange, but one should take into account that the wind shear $w\tau$ must increase together with $D$ to preserve the validity of the long-exposure regime.

An entirely different situation arises in the case of perpendicular wind, when the magnification of the colour scintillation due to wind depends only on the wavelengths, as in the case of (\ref{eq:rel_ampl}):
\begin{equation}
\frac{\Delta\tilde s^2_d}{\tilde s^2_d} = 3.1\,y^2\frac{\mathcal I_4(\lambda_1/\lambda_2)}{\mathcal I_2(\lambda_1/\lambda_2)} \approx 800\,y^2.
\end{equation}
It is seen that the intensity of colour scintillation is tripled even at $y = 0.05$ or $M_z \approx 1.05$, increasing by $\approx 30$ times at $M_z = 1.3$.

The above estimates are confirmed by the exact calculations, the results of which are shown in Fig.~\ref{fig:real_disp_l}. The functions $U^{\prime}_{\Delta}(z)$ are calculated for telescopes of 1~m and 2.5~m diameter with central obscuration 0.3, polychromatic $B$ and $R$ bands, and the exact dependence of linear atmospheric dispersion $x$ on the altitude.

A large difference between these two situations is seen. The effect of the parallel wind at $M_z = 1.5$ is much less than the effect of the transverse wind at $M_z = 1.05$. Parallel WFs are calculated for a small wind shear $w\tau = 5$~m; at larger wind shears the contribution of the atmospheric dispersion will only become smaller. For the perpendicular wind, the functions do not depend on the wind shear, although of course, the scintillation index will decrease in proportion to $\sim 1/w\tau$ in any case.

In analogy with (\ref{eq:final0}), we can write:
\begin{equation}
\frac{\Delta\tilde s^2_d + \tilde s^2_d}{ \tilde s^2} \approx (0.0003 + 0.012\,M_z(M_z^2-1))\,D^{-5/3}.
\label{eq:finallarge}
\end{equation}

For this approximation, a worst-case scenario is adopted: the wind is perpendicular to the beam displacement due to the atmospheric dispersion.

In the formula (\ref{eq:finallarge}) as well as in the expression (\ref{eq:uddz_per}), the wind speed is not present explicitly. It enters through the condition $w\tau \gg D$. This means that for any small perpendicular component of the wind but rather long exposure, the worst case situation may be encountered.

\section{Discussion}

As an example, the magnitudes of the normal and the colour scintillations are presented in Table.~\ref{tab:example} for different situations. They are calculated using the exact WFs (Fig.~\ref{fig:real_disp_s},~\ref{fig:real_disp_l}) and a real turbulence profile from MASS measurements at Mount Shatdzhatmaz \citep{kgo2010}. One specific turbulence profile was chosen for which the atmospheric moment $M_2$ corresponded to its median. In the long-exposure regime, the values $s^2 w\tau$ (see Sect.~\ref{sec:wind_zero}) are given, from which the scintillation indices can be obtained by dividing those values by $w\tau$.

\begin{table}
\caption{Estimation of the normal (N) and the colour (C) scintillation indices in the $B$ and $R$ spectral bands for 1~m and 2.5~m telescopes in the short (SE) and long (LE) exposure regimes. For the LE regime, two situations are considered: with the wind perpendicular (LE${}_{\perp}$) and parallel (LE${}_{\parallel}$) to direction of the atmospheric dispersion. For clarity, the amplitudes $\sqrt{s^2}$ are written in millimagnitudes in the SE case and in millimagnitudes per meter in the LE case.
\label{tab:example}}
\begin{tabular}{llrrrrrr}
\hline
 & &\multicolumn{2}{c}{SE} & \multicolumn{2}{c}{LE${}_{\perp}$}& \multicolumn{2}{c}{LE${}_{\parallel}$} \\
 & $M_z$ & 1~m & 2.5~m & 1~m & 2.5~m & 1~m & 2.5~m \\
\hline
N & 1.0 & 14.48 & 5.21 & 10.85 & 5.91 & 10.85 & 5.91 \\
C & $1.0$ & 1.26 & 0.33 & 0.23 & 0.06 & 0.23 & --- \\
C & $1.05$ & --- & --- & 0.43 & 0.12 & --- & --- \\
C & $1.1$ & 1.64 & 0.43 & 0.57 & 0.16 & --- & --- \\
C & $1.2$ & 2.08 & 0.54 & 0.88 & 0.25 & --- & --- \\
C & $1.3$ & 2.56 & 0.67 & 1.18 & 0.33 & 0.25 & 0.07 \\
C & $1.5$ & --- & --- & --- & --- & 0.27 & 0.08 \\
\hline
\end{tabular}
\end{table}

These estimates confirm that colour scintillation is much less than normal scintillation when observing near the zenith. This is especially noticeable in the long-exposure regime typical for conventional photometry. The effect of atmospheric dispersion favours observations near the zenith, but colour scintillation is reduced at large telescopes, allowing observations at larger air mass.

Obviously, the diameter of telescope aperture is also important for another reason. The gain from potential scintillation noise suppression by multicolour photometry is achievable only when the photon noise is low enough. Estimation of the limiting object is straightforward. Only two points will be emphasised here.

The ratio of normal scintillation to the photon noise depend on the exposure time. In the short-exposure regime, the scintillation noise is $\propto D^{-7/6}$ and decreases slightly faster than the photon noise $\propto D^{-1}$ with increasing telescope diameter $D$. In the long exposure regime, the scintillation decreases as $\propto D^{-2/3}$, much slower than the photon noise. Therefore, in this situation, the scintillation noise becomes dominant at large telescope apertures.

The colour scintillation, on the other hand, always falls faster ($\propto D^{-3/2}$) than the photon noise. Consequently, when the telescope diameter increases, the photon noise becomes, sooner or later, the main limitation of the measurement accuracy. So, we can take advantage of the small colour scintillation only by choosing a bright enough astronomical object.

A similar situation exists in the case of photometry from space. In the space missions CoRoT \citep{Corot2009} and Kepler \citep{Kepler2010}, the photon noise determines the fundamental limit of photometric accuracy. However, the aperture of these instruments (CoRoT: 0.27~m, Kepler: 0.95~m) are much smaller than the sizes of ground-based telescopes. This leads to a potential advantage of ground-based observations: it is possible to achieve a better temporal resolution or to measure fainter objects.

In the above example, the standard photometric $B$ and $R$ wide bands, often used for studies of the micro-variability, were adopted. Their large difference in wavelength helps to detect the colour variability of an astronomical object in the presence of spectral differences in light emitting processes. Asymptotic dependencies derived here allow us to obtain estimates for the opposite case of close and narrow photometric bands. For instance, to determine the variability in the spectral lines of radiation. When wavelengths difference is about $100 - 500$\,\AA, the effect of two decorrelation mechanisms (atmospheric dispersion and wind drift) is 1-2 orders of magnitude smaller than for the $B$ and $R$ bands.

\section{Conclusion}

Although a large telescope aperture is comparable to the turbulence outer scale, we did not assess its potential impact on the obtained asymptotic dependencies. Probably, such accounting for can slightly change the dependence on the telescope diameter, but we do not expect a large effect because scintillation is produced mainly by the first and subsequent peaks of the Fresnel spectral filter.

Also, implications of a non-Kolmogorov turbulence spatial spectrum were not analysed. The preliminary estimate of this effect was made in \citep{multi2011} for the case of small apertures, where it was shown that a flatter spectrum weakens correlation of the scintillation in different photometric bands (increases colour scintillation). This is clear, because such spectrum leads to a raise of the contribution of the high frequencies which are transmitted by the differential Fresnel filter.

We summarise briefly the main results and conclusions obtained in this paper under the assumption of the Kolmogorov spectrum of optical turbulence:
\begin{description}
\item[--] The colour scintillation was considered. It is produced by the wavelength dependence of the light diffraction in the turbulent atmosphere.
\item[--] The spatio-chromatic scintillation caused by the atmospheric dispersion was analysed in addition.
\item[--] The asymptotic dependencies were obtained to estimate the colour scintillation as a function of the telescope diameter.
\item[--] The dependencies of the colour scintillation power on the turbulence altitude were determined (WFs for colour scintillation).
\item[--] The colour scintillation is generated by highest turbulence and grows faster with the air mass than the normal scintillation.
\item[--] Analysis of the behaviour of the scintillation was performed both for short exposures and for long exposures. The last is typical in conventional photometry.
\item[--] In the short exposure regime, the achromatization of the  scintillation occurs slowly. To halve  the colour scintillation, the telescope diameter must increase by 8 times.
\item[--] In the long exposure regime, the achromatization is happening fast enough: when the telescope diameter increases by 2.3 times, the relative power of the colour scintillation is halved.
\item[--] It was found that the combined effect of wind smoothing and atmospheric dispersion leads to a strong dependence of intensity of the colour scintillation on the wind direction in the upper atmosphere.
\item[--] With an increase in the diameter of telescope, the colour scintillation is falling faster than the photon noise, so eventually the limiting accuracy is determined by the photon noise.
\end{description}

Therefore, when observing on large telescopes, the scintillation in two different photometric bands is strongly correlated. The colour scintillation is much less than the normal one and in the circumstances discussed above, the practical application of simultaneous multicolour measurements will provide distinct advantages in photometric accuracy.

\section{Acknowledgements}

The author thanks A.\,Tokovinin for an interesting discussion of this work, and B.\,Safonov for critical comments. He is also grateful to the referee for very interesting comments on the additional aspects of the problem of accurate photometry.

\bibliography{reference_list}
\bibliographystyle{mn2e}

\label{lastpage}

\end{document}